\documentclass[a4paper,twocolumn,showpacs,preprintnumbers,amsmath,amssymb,nofootinbib]{revtex4}
\input psfig.sty 
\usepackage{graphicx}

\begin{document}

\title{Cosmography and cosmic acceleration}

\author{J. C. Carvalho$^{1,2}$\footnote{E-mail: carvalho@dfte.ufrn.br}}

\author{J. S. Alcaniz$^2$\footnote{E-mail: alcaniz@on.br}}

\address{$^1$Departamento de F\'isica, Universidade Federal do Rio Grande do Norte, 59072-970, Natal - RN, Brasil}

\address{$^2$Departamento de Astronomia, Observat\'orio Nacional, 20921-400, Rio de Janeiro - RJ, Brasil}

\date{\today}

\begin{abstract}
We investigate the prospects for determining the accelerating history of the Universe from upcoming measurements of the expansion rate $H(z)$. In our analyses, we use Monte Carlo simulations based on $w$CDM models to generate samples with different characteristics and calculate the evolution of the deceleration parameter $q(z)$. We show that a cosmographic (and, therefore, model-independent) evidence for cosmic acceleration ($q(z<z_t) < 0$, where $z_t$ is the transition redshift) will only be possible with an accuracy in $H(z)$ data greater than the expected in current planned surveys. A brief discussion about the prospects for reconstructing the dark energy equation of state from the  parameters $H(z)$ and $q(z)$ is also included.

\end{abstract}


\maketitle

\section{Introduction}
\label{intro}

The determination of cosmographic parameters, such as $H_0$ and $q_0$, has a long and interesting history in Cosmology~(see, e.g., \cite{sand}). In particular, the evolution of such parameters provides a unique and direct method to map the expansion history of the Universe in a model-independent way. Since all evidence we have so far for the current cosmic acceleration are indirect~\cite{review}, extracting the evolution of these two parameters from future redshift surveys constitutes one of the major challenges in observational cosmology\footnote{Due to the multiple integrals that relate cosmological parameters to cosmological distances, direct determinations of $H(z)$ may also reduce the so-called smearing effect  that makes constraining the dark energy equation of state (EoS) $w$ extremely difficult~\cite{steinhardt}.}.

In this paper, we investigate how well cosmography may provide a model-independent way to check the reality of cosmic acceleration. Specifically, we study the evolution of the deceleration parameter from the $H(z)$ data which are to become available by some planned projects. To this end, we performed a Monte Carlo (MC) simulation based on $w$CDM models (of which the $\Lambda$CDM model is a special case with $w \equiv p/\rho = -1$, where $p$ stands for the dark energy pressure and $\rho$ for its energy density). We generated three samples of $H(z)$ with increasing accuracy and derived $q(z)$ from numerical differentiation. We show that a comographic evidence for cosmic acceleration will be possible only with a great  (maybe out of the perspective of current planned surveys) accuracy in $H(z)$ data. It is worth mentioning that the determination of $q(z)$ will also allow us to estimate the transition redshift, $z_t$, which corresponds to the epoch when the Hubble expansion switched from a decelerating to an accelerating phase. Since $q(z) \propto w(z)H(z)^2/f(z,w)$, the same technique can also be used to map the recent past evolution of $w(z)$ in a given model of dark energy.

\begin{figure*} \label{Fig1}
\centerline{\psfig{figure=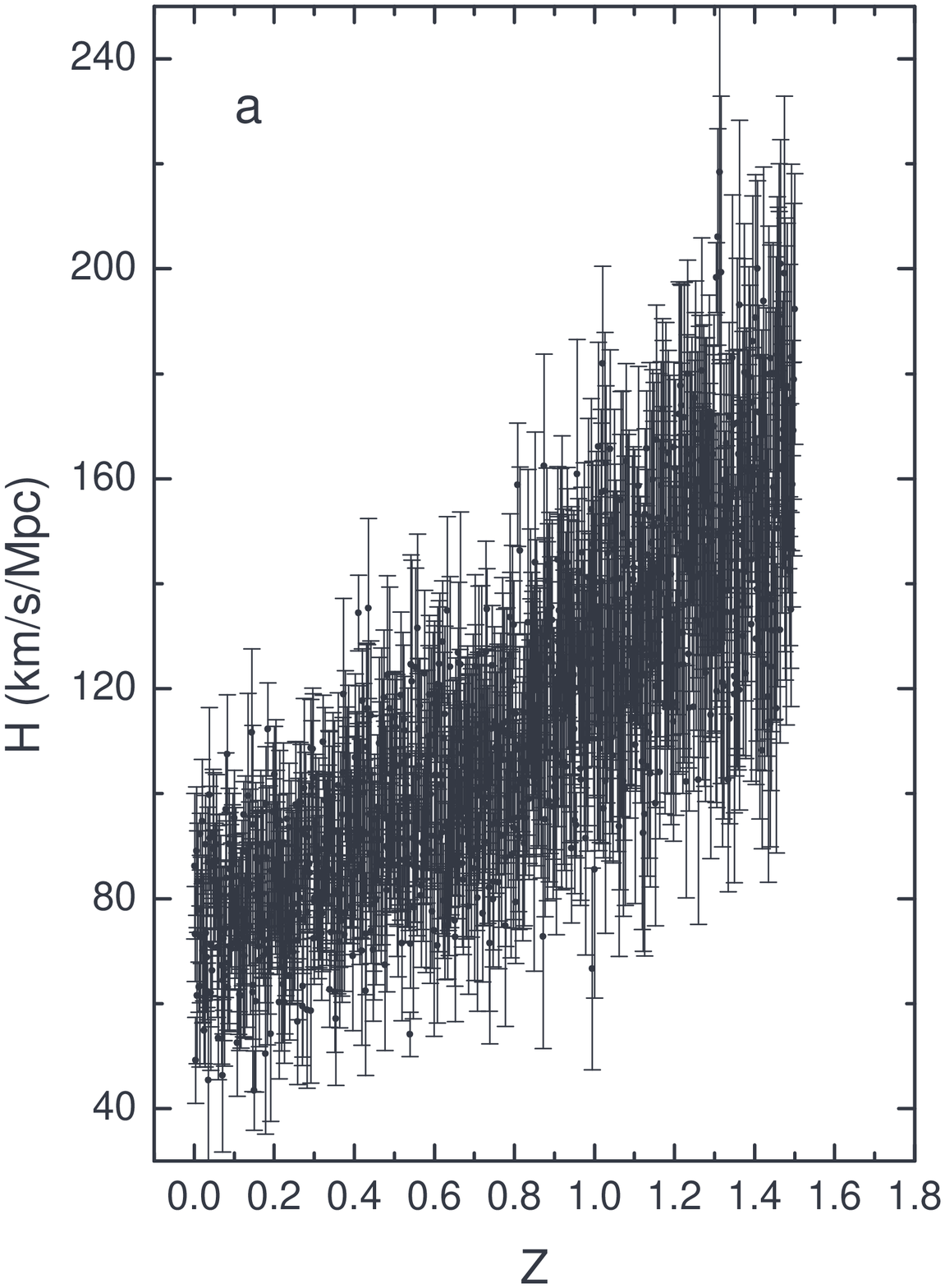,width=2.3truein,height=1.9truein}
                   \psfig{figure=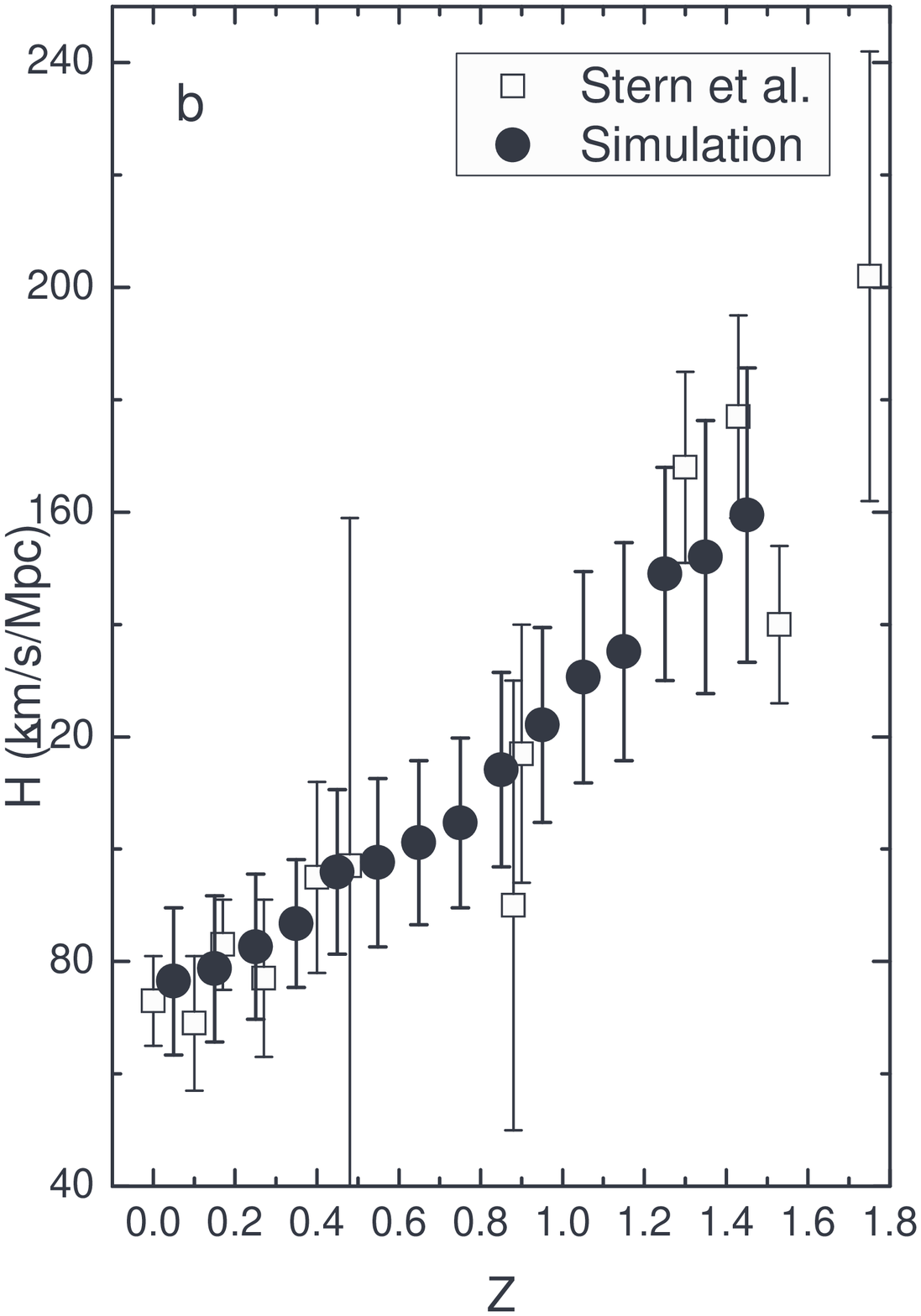,width=2.3truein,height=1.9truein}
                    \psfig{figure=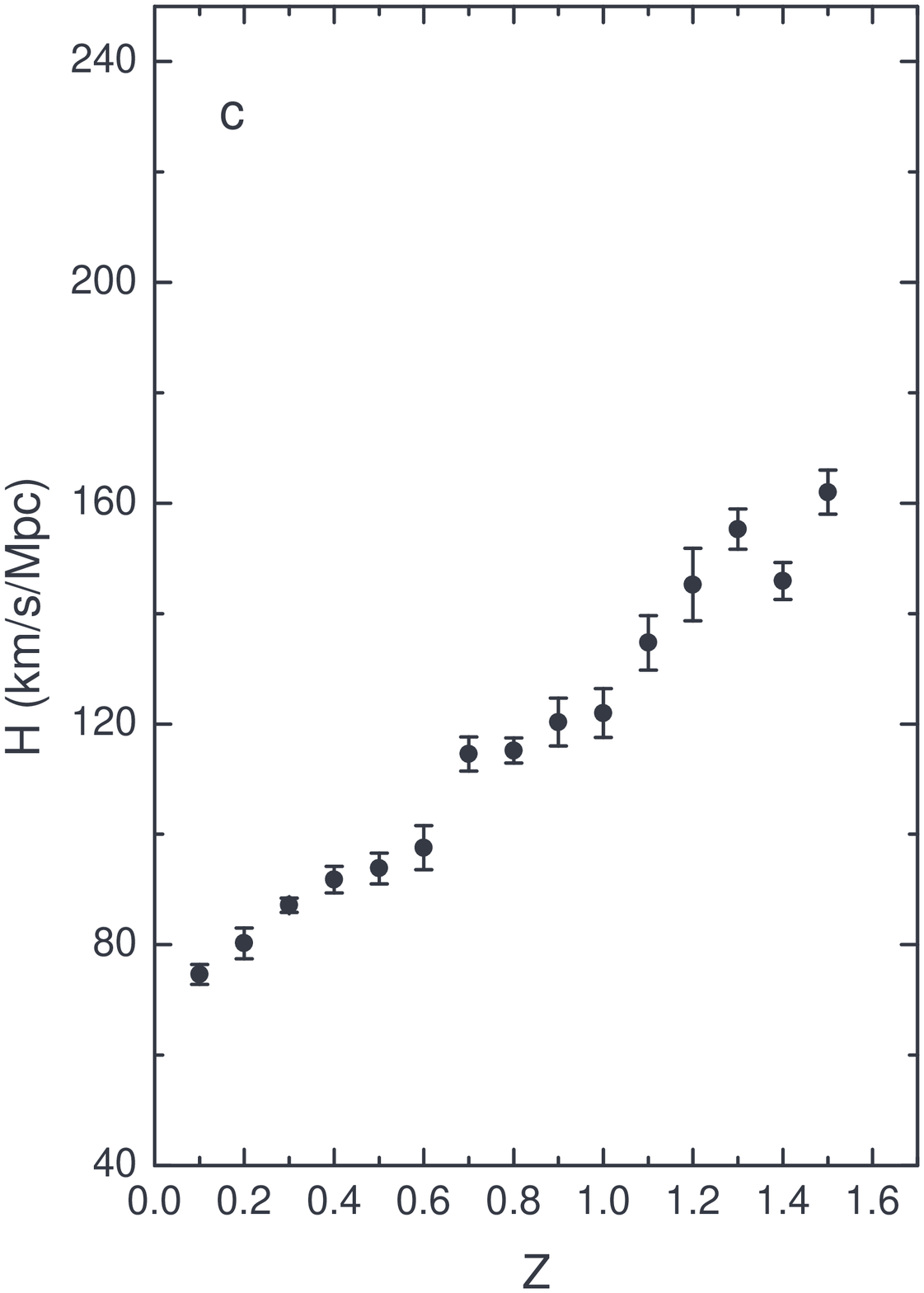,width=2.3truein,height=1.9truein}}
\caption{{{a)}} A MC realization of 1000 simulated values of the Hubble parameter with 15\% accuracy based on Ref.~\cite{svj}.  {{b)}} Simulated values of $H(z)$ shown in the previous panel averaged over bins with width $\Delta z=0.1$. Squares represent current $H(z)$ measurements of Ref.~\cite{stern}. {{c)}} Similar to Panel (a) for the 3\% projection made in Ref.~\cite{crawford}.}
\end{figure*}

\section{Direct observation of $H(z)$}

Recently, it has been shown that Luminous Red Galaxies (LRG's) can provide us with direct measurements of the expansion rate $H(z)$~\cite{jimenez} (see also \cite{ma} for a recent review on $H(z)$ measurements from different techniques)\footnote{Direct measurements of $H(z)$ at different redshifts will also be possible through measurements of the line-of-sight or radial component of baryonic acoustic oscillations (BAO) from large redshift surveys with redshift precision of the order of $0.003(1+z)$ (see, e.g., \cite{pau}).}. This can be done by calculating the derivative of cosmic time with respect to redshift
\begin{equation}
H(z) = -\frac{1}{(1+z)}\frac{dz}{dt}\;,
\end{equation}
from measurements of age difference between two passively evolving galaxies at different $z$. Simon, Verde and Jimenez~\cite{svj} (SVJ) have demonstrated the feasibility of directly measuring $H(z)$ using this differential age method and have provided 9 determinations of the expansion rate at $z \neq 0$. More recently, Stern {\it{et al.}} \cite{stern} have updated SVJ sample to 11 estimates of $H(z)$ lying in the redshift interval $0.1 \leq z \leq 1.75$. New age-redshift datasets for different galaxy velocity dispersion groups (7 LRG's sample) have also been made available by the SDSS collaboration~\cite{sdss7LRG}.

Ref.~\cite{svj} have also pointed out that in the near future,  the Atacama Cosmology Telescope (ACT)\footnote{http://www.physics.princeton.edu/act/} is expected to provide observations of over 500 galaxy cluster up to $z \lesssim 1.5$. This, together with spectra to be acquired in other telescopes in Chile and Southern African Large Telescope (SALT) in South Africa, will provide a sample of more than 2000 passively evolving galaxies in the redshift range $0<z<1.5$. From these observations, it will be possible to determine $\sim 1000$ values of the Hubble parameter at a $15\%$ accuracy level if the ages of the galaxies are estimated with  $10\%$ error.

Following a similar approach, Crawford {\it{et al.}}~\cite{crawford} examined the observational requirements to estimate $H(z)$ to a given precision. In their simulation, observations of LRG's are made at two redshifts, namely, $z = 0.32 $ and $z = 0.51$ (average $0.42$) and they supposed that uncertainty on ages of individuals galaxies lies in the range  $0.05 - 2$ Gyr. They estimated that the uncertainty on the mean ages will be 0.10, 0.05 and $0.03$ Gyr and that Hubble parameter can be measured with a precision of $10\%$, 5\% and $3\%$. A study was also made on the constraints upon the exposure time per galaxy and number of galaxies observed. They concluded that with a total time of 17, 72 and 184 hours, observing 80, 327 and 840 galaxies on the Southern African Large Telescope, will make possible to recover $H(z)$ to 10\%, 5\% and $3\%$ accuracy, respectively~\cite{crawford} (private communication).

\begin{figure*} \label{Fig3}
\centerline{\psfig{figure=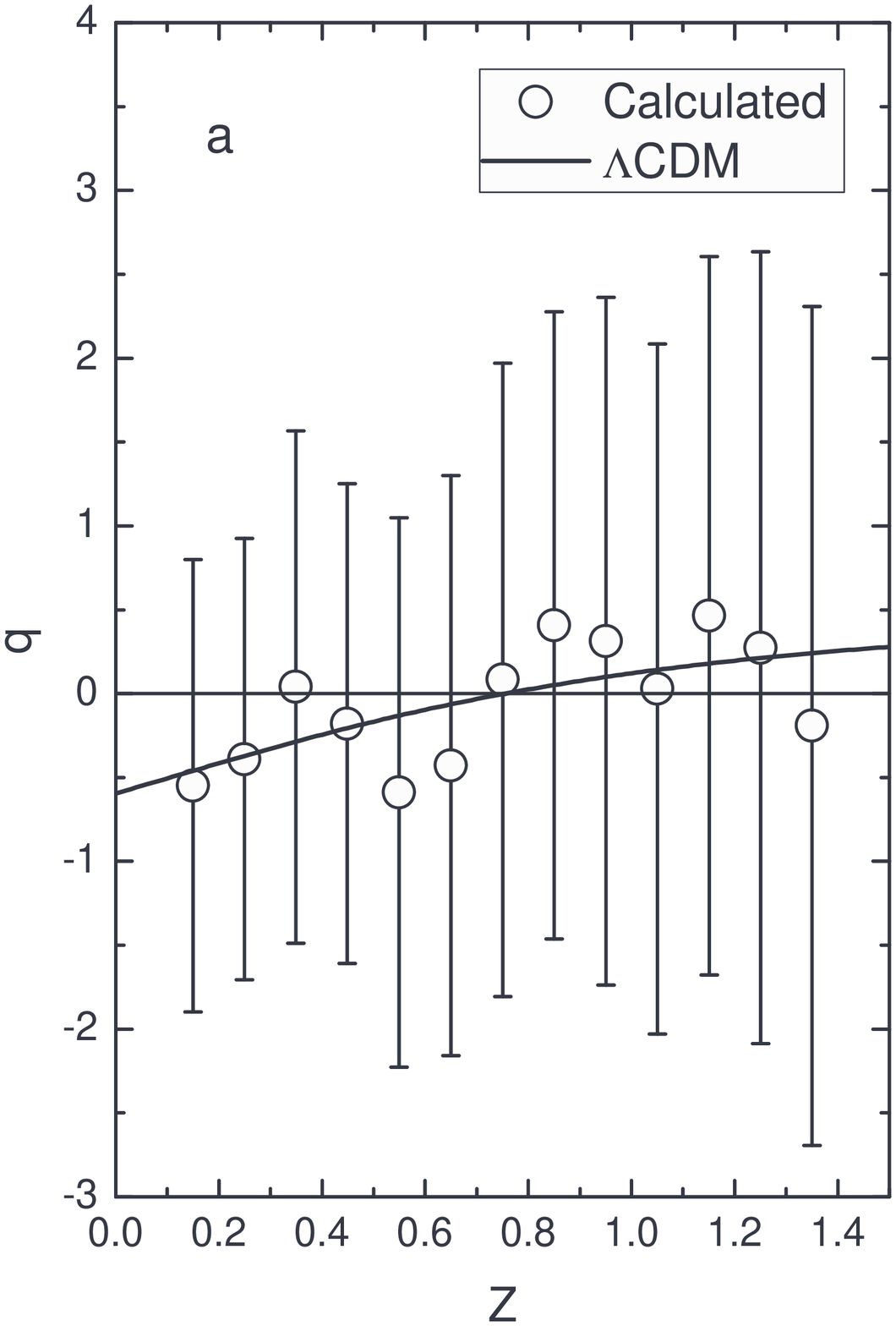,width=1.8truein,height=1.8truein}
\psfig{figure=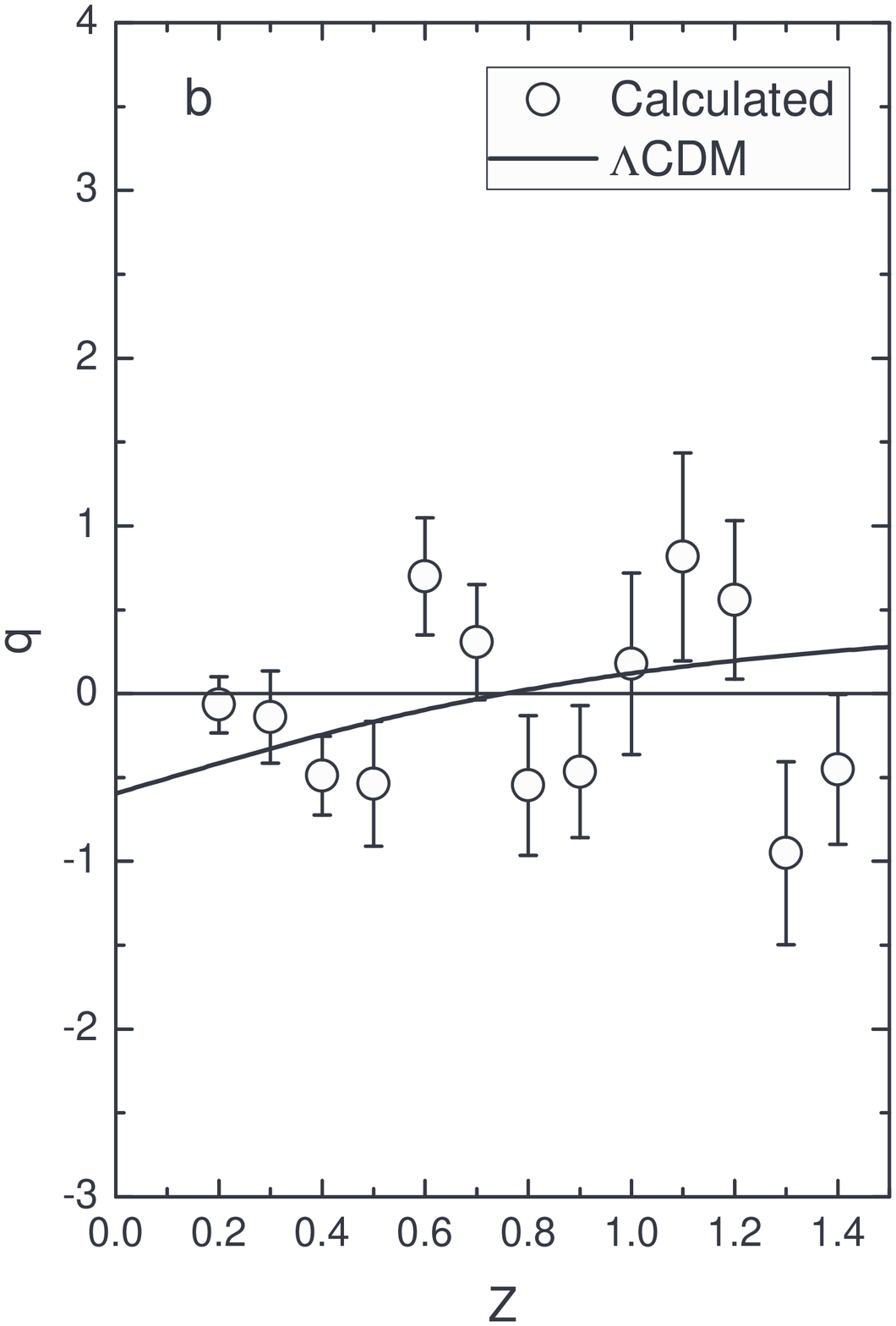,width=1.8truein,height=1.8truein}
\psfig{figure=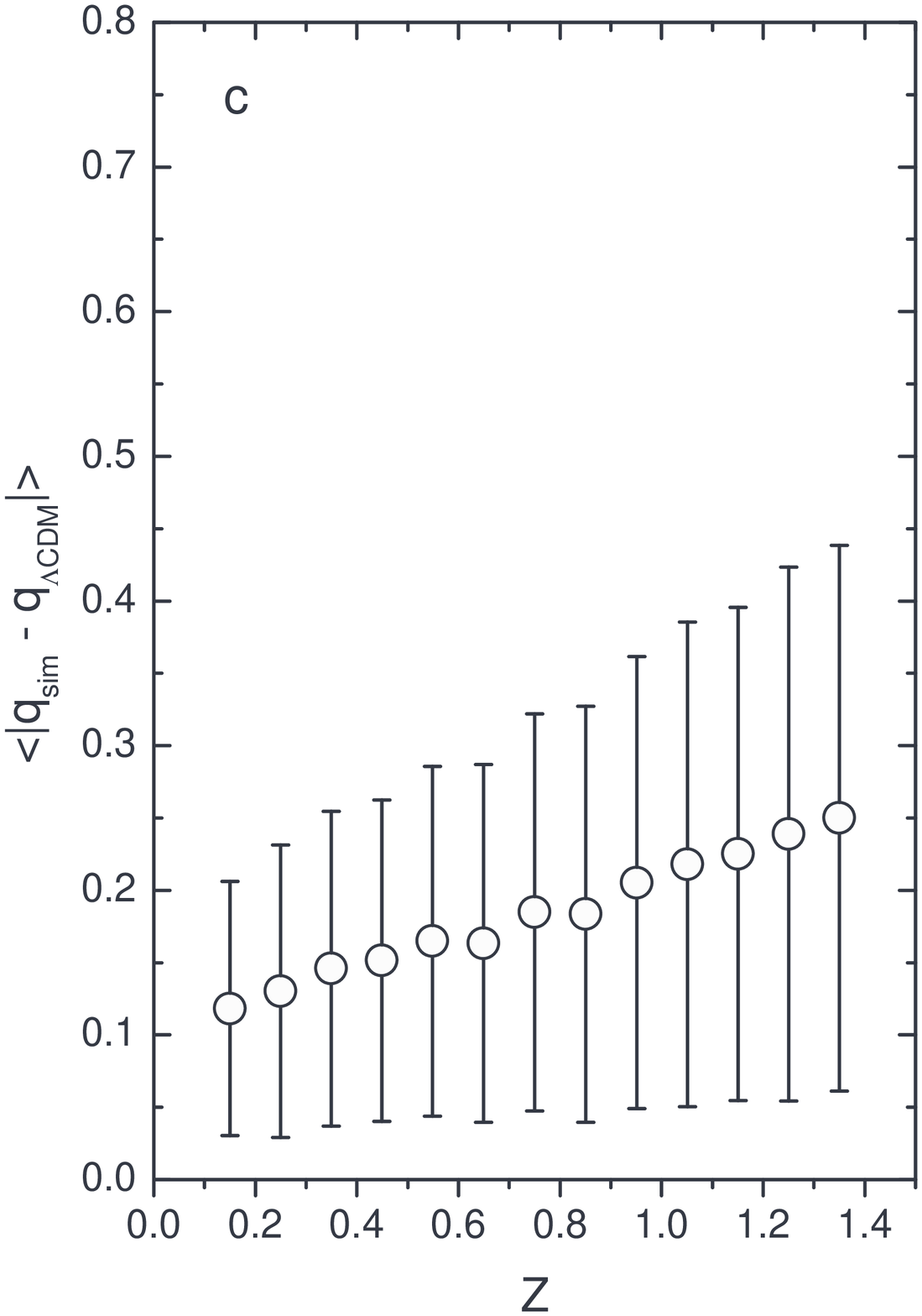,width=1.8truein,height=1.8truein}
\psfig{figure=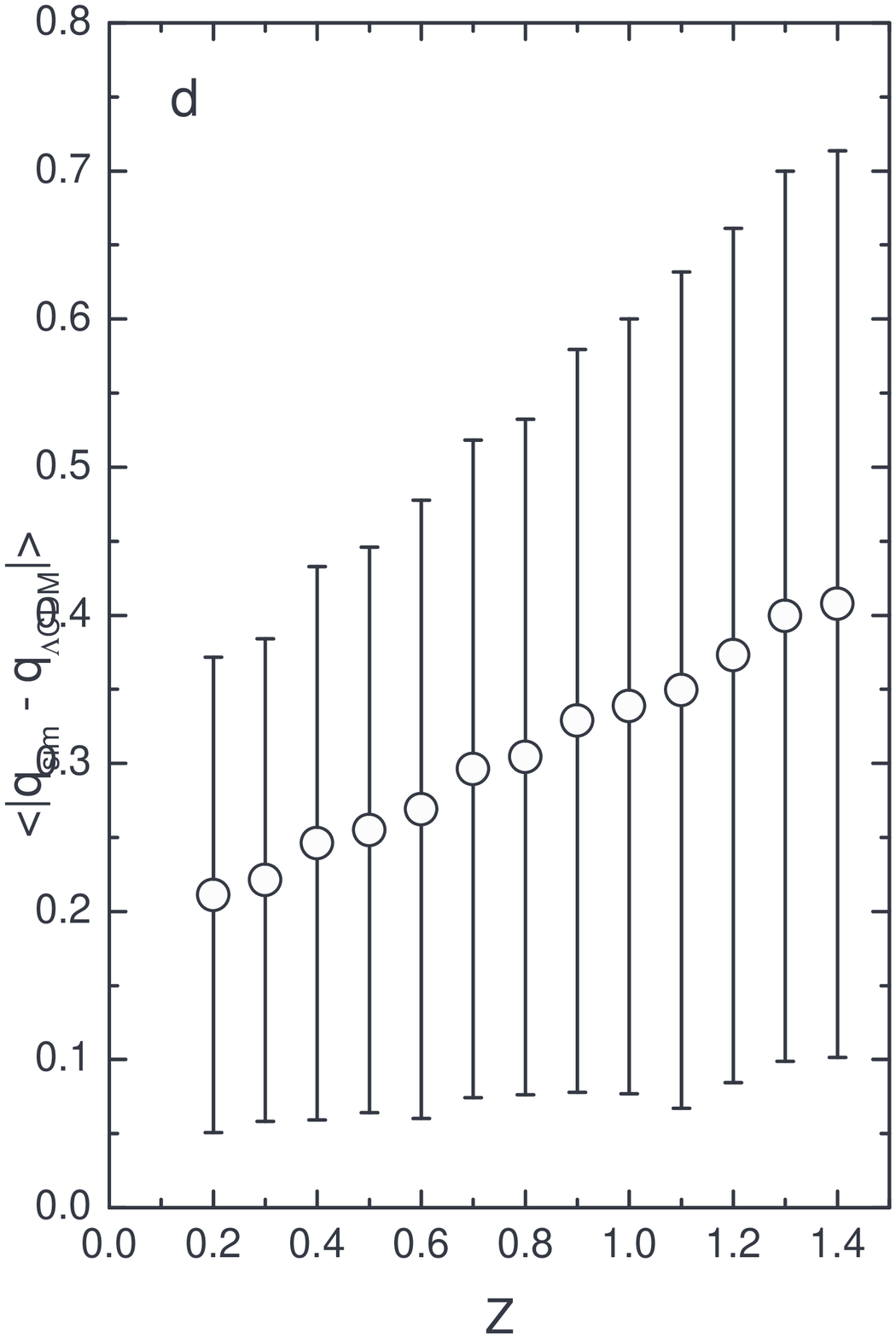,width=1.8truein,height=1.8truein}}
\caption{The evolution of $q(z)$ derived from the $H(z)$  points displayed in Figure 1. Panel a corresponds to the 1000 binned values of Fig. 1b whereas Panel b to the 15 $H(z)$ data points at 3\% accuracy shown in Fig. 1b. c) The difference between the simulated and the background model values of $q(z)$ for 1000 MC realization and 15\% accuracy in $H(z)$ observations. d) The same as in Panel c for the 15 $H(z)$ data points at 3\% accuracy shown in Fig. 1b.}
\end{figure*}

\begin{figure*} \label{Fig2}
\centerline{\psfig{figure=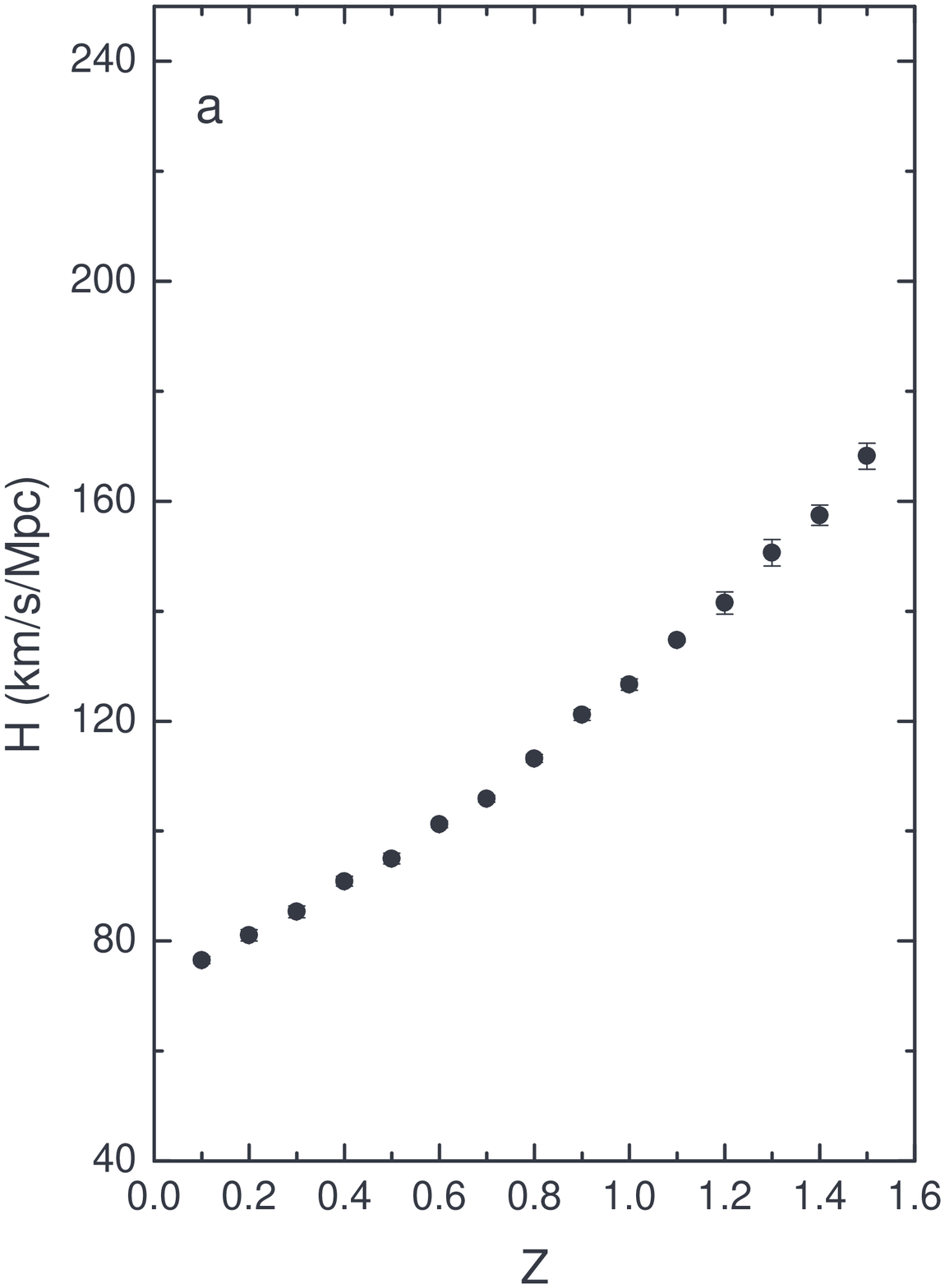,width=2.3truein,height=1.8truein}
\psfig{figure=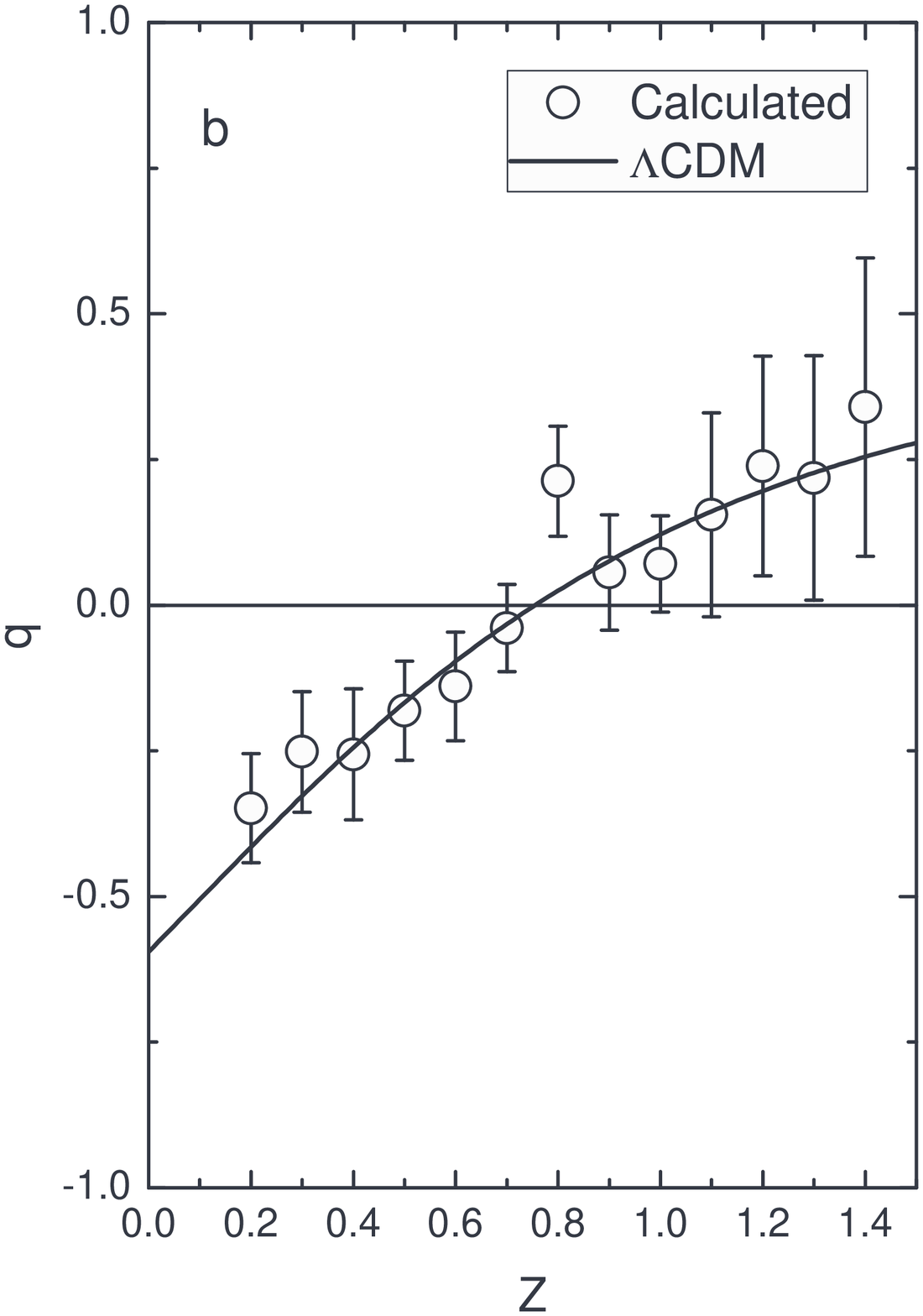,width=2.3truein,height=1.8truein}
 \psfig{figure=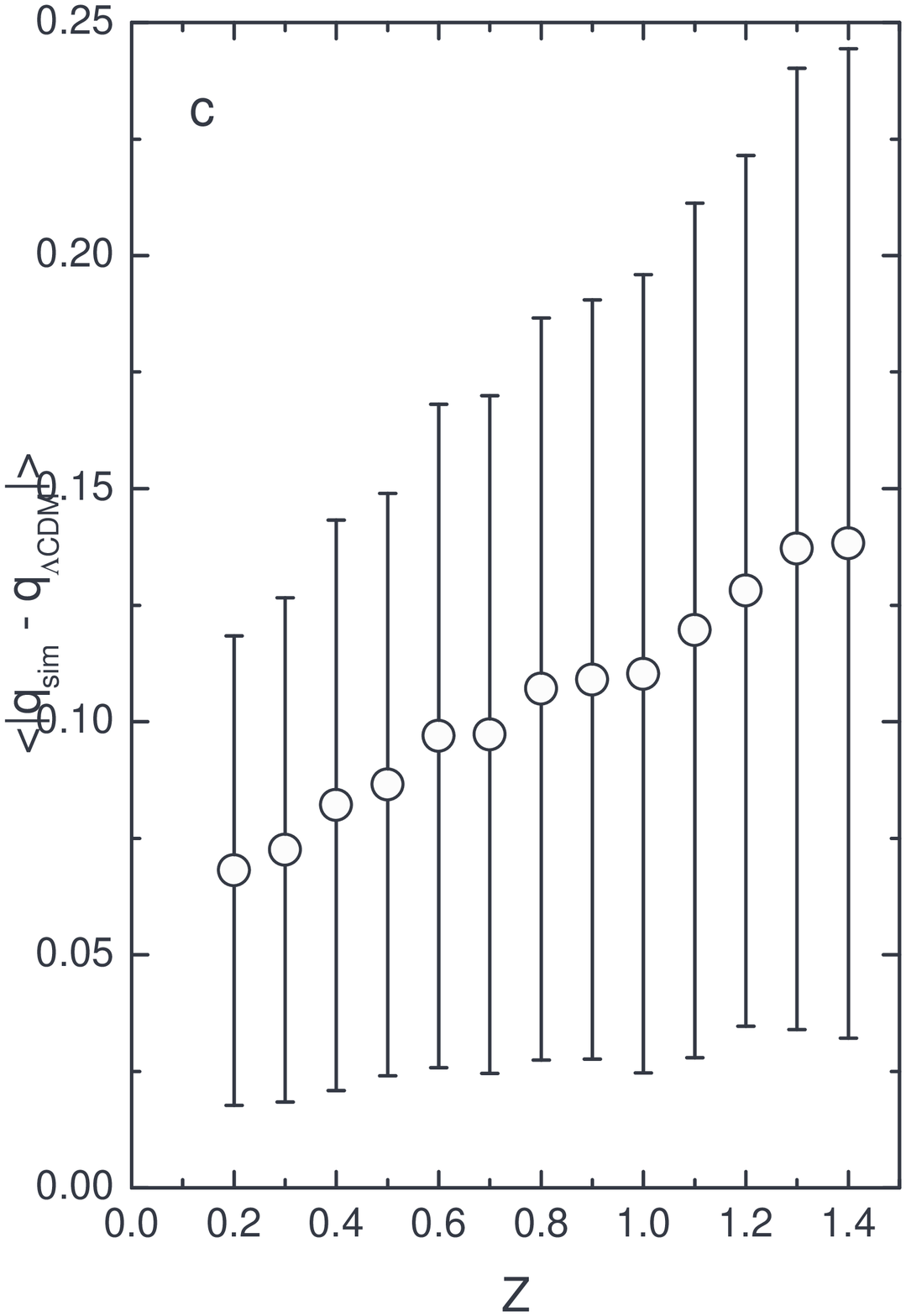,width=2.3truein,height=1.8truein}}
\caption{a) The results of our simulations following~\cite{crawford} with a precision of 1\% in the $H(z)$ observations. {{b)}} The evolution of $q(z)$ for the MC realization shown in Panel a. With this precision in the $H(z)$ measurements, a cosmographic detection of of the current cosmic acceleration can be obtained, as can also be seen from Panel c.}
\end{figure*}

\section{Simulated datasets and $q(z)$}


In our analyses, we perform Monte Carlo simulations which provide us with samples of $H(z)$ based on the flat $\Lambda$CDM model. For each "measurement", we assume a Gaussian distribution of $H(z)$ centered at the value predicted by a flat $w$CDM with $w = -1$ with a standard deviation corresponding to the percentage accuracy predicted in future experiments. We shall examine two cases based on the projections made in Refs.~\cite{svj} and \cite{crawford}. For the former case, we simulated 1000 data points for $H(z)$, uniformly distributed between $z=0$ and $z=1.5$ with $15\%$ precision. In the latter case, we simulated 15 measurements of the Hubble parameter between redshifts 0 and 1.5 and equally spaced ($\Delta z=0.1$). We take the errors estimated in Ref.~\cite{crawford} when recovering $H(z)$ from LRG's ages, namely 10\%, 5\% and $3\%$. Note that in the last case (the one explored throughout this paper), one needs 8400 LRG's in the whole interval $z=0.1 - z=1.5$. In all our simulations, we have adopted  $H_0=74.2$ km/s/Mpc, which corresponds to the central value given in Ref.~ \cite{riess} based on differential measurements of Cepheids variable observations and $\Omega_m=0.27$, as given by current CMB measurements~\cite{wmap}.

Figure 1a shows one realization of 1000 simulated values of $H(z)$ with $15\%$ accuracy according to Ref.~\cite{svj} for $z$ in the range ($0.0 - 1.5$). The binned data points of this realization (with a bin width $\Delta z=0.1$) are shown in Figure 1b. The error bars correspond to the standard deviation around the mean of individual values of $H(z)$ within each bin, which contains approximately 67 galaxies. For comparison, the observed values given in \cite{stern} are also shown. In Figure 1c we show the results of one MC realization taking the errors estimated in Ref.~\cite{crawford} with the resulting $H(z)$ data points at $3\%$ accuracy.


From the above $H(z)$ simulated data, one may derive evolution of the deceleration parameter $q(z)$, defined as
\begin{equation}
\label{q1}
q(z)={1 \over H(z)} \left[ {dH(z)\over dz}\right](1+z)-1.
\end{equation}
In order to do that one needs to compute the derivative of the $H(z)$ with respect to the redshift. However, while integration tends to smooth out data fluctuations, numerical differentiation tends to magnify errors and the scatter of the points. Thus, depending on the degree of scatter of $H(z)$ measurements, the gradient estimation may be useless.
For equally spaced points, the derivative is calculated using finite difference approximation, that is, $dH(z_i)/dt\simeq [H(z_{i+1})-H(z_{i-1})]/2\Delta z$, where $\Delta z=z_i-z_{i-1}$. We then substitute this into (\ref{q1}) and calculate $q(z_i)$ and the associated uncertainty by using the standard error propagation method.

For our first simulation, shown in Figures 1a and 1b, we use the binned points since the large scatter of the raw data makes them useless. The resulting values of the deceleration parameter from the MC realization of Figure 1b are shown in Figure 2a. Although the points lie near the predicted curve of the $\Lambda$CDM model (black curve), we cannot expect a definitive evidence for a recent cosmic acceleration from these simulated $H(z)$ sample with $15\%$ accuracy. Note that the same is also true even when the evolution of $q(z)$ is derived from the 15 data points with $3\%$ accuracy following \cite{crawford} (Fig. 2b). In this case, although the error of individual points is smaller, the determination of the deceleration parameter is less precise with considerable scattering of points and error bars allowing both a decelerating and accelerating universe today.

In order to quantify the above results, we also performed 1000 MC realizations to calculate the mean of the absolute values of the difference between the calculated and the background model values of $q_{\Lambda \rm{CDM}}(z_i)$, that is,
\begin{equation}
 <\Delta q> = {1\over N} \sum_{i=1}^N |q_{sim}(z_i) - q_{\Lambda \rm{CDM}}(z_i)|.
\end{equation}
We plot this quantity for both cases discussed above in Figures 2c (1000 binned points) and 2d (15 points), with the error bars representing the standard deviation from the mean. As we see, $<\Delta q>$ has a weak dependence on redshift for the first case and in the interval $0.1<z<1.4$ it lies in the range $\sim 0.1 - 0.25$, although the uncertainty is relatively large. From Figure 2d we see that the departure of the calculated $q_{sim}(z_i)$ from the expected value given by the model is  larger than in former case, with $<\Delta q>$ varying from 0.2 to 0.4. This clearly reflects the large scattering of points shown in Figure 2b.

In view of the above results, we also speculated about how precise should future observations be in order to provide a clear evidence for cosmic acceleration from cosmography. We did that for the method proposed in \cite{crawford} and found that it will give very good results if the accuracy is improved to as much as $1\%$. This is illustrated in Figure 3. In this case, the calculated value of $q_{sim}(z_i)$ is very close to the $\Lambda$CDM model predictions in almost the entire range of $z$ and the deviation from the expected value is as low as 0.07 and always $< 0.15$. Naturally, the observational requirements for this case would be much more stringent than the previous ones, with the need of thousand galaxies and more observational time as well.

\section{Equation-of-state parameter}

By considering a dark-energy dominated universe, a step further in the above discussion concerns the reconstruction of its equation-of-state parameter $w$ from measurements of $H(z)$. In order to be consistent with the simulations presented earlier,  we consider accelerating $w$CDM models with $w$ lying in the interval [-1/3, -3/2]. In this context, Eq. (\ref{q1}) can be rewritten as
\begin{equation} \label{wz}
w(z) = {2q(z)-1\over {3\Omega_w^0 (1+z)^{3(1+w)}}}{\left(H \over H_0\right)^2}\;,
\end{equation}
where  $\Omega_w^0$ is the current fractional contribution of dark energy to the critical density. Note that such procedure does not involve several integrations of $w(z)$, as in cosmological tests based on age or distance measurements and, therefore, may reduce the smearing effect on $w(z)$ determinations discussed earlier (see also \cite{jimenez} for a different approach).

Figure 4 shows the evolution of $w$ with  redshift  calculated from the simulated measurements of $H(z)$ using the method proposed of Ref.~\cite{crawford} with $1\%$ accuracy. Although the input model ($w = -1$) is fairly recovered, we note that  the estimated errors  of the dark energy EoS  ($\delta w$) increase considerably at the upper end of the redshift range. This can be more easily understood after solving Eq. (\ref{wz}) for $w$ and calculating $\delta w$, which depends strongly on a competition between  the different terms of Eq. (\ref{wz}) at the $z$ interval considered. For the sake of completeness, we also performed the same analysis for the other two cases (with 15\% and 3\% accuracy in $H(z)$) discussed earlier. In both,  the scattering of the points is large enough to hamper any definitive conclusion on the expected evolution of the cosmic equation of state.

\section{Conclusions}

\begin{figure}[t] \label{Fig3}
\centerline{\psfig{figure=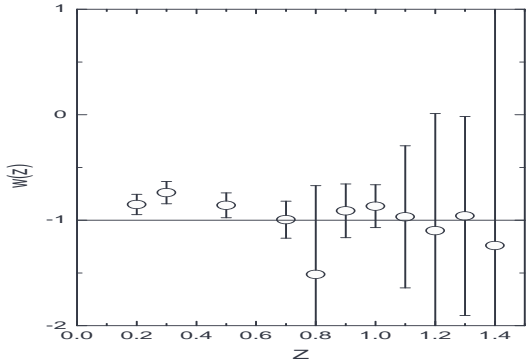,width=3.4truein,height=2.3truein}}
\caption{Evolution of $w$ with redshift for 1\% accuracy in $H(z)$ observations following the method of Ref.~\cite{crawford}. Although the background model is fairly recovered, the uncertainties increase considerably at the upper end of the redshift interval considered.}
\end{figure}

Cosmography explores the possibility of extracting the maximum amount of information from cosmological measurements, as well as the assumption that the universe can be modeled by the Friedmann-Robertson-Walker line element without assuming any dynamical theory to describe it. In this paper, differently from many analyses that study the current phase of cosmic evolution in the context of a given cosmological model, we have investigated the prospects for a cosmographic mapping of cosmic acceleration using two different projections of age determinations of passively evolving galaxies recently discussed in the literature~\cite{svj,crawford}. We have found that a model-independent check of the reality of cosmic acceleration requires very accurated measurements of the expasion rate ($\sim 1\%$), which seems to be far from the perspectives of current planned surveys. As a step further in this analysis, we have also used the direct relation between $q(z)$ and $w(z)$ to study possible constraints on the dark energy EoS from future observations (Fig. 4).

Finally, it is also worth emphasizing that the method discussed here can be applied regardless of the technique used to obtain $H(z)$ measurements. In this regard, we note that $H(z)$ measurements from radial BAO methods (despite the systematic errors introduced mainly by distortion effects) surpasses the age method in precision and can extend the $H(z)$ measurements into deeper redshift ranges~\cite{pau} (see also \cite{lyman} for expected measurements of expansion rate at $z \leq 2.5$ by observing the Lyman-forest absorption spectra of high-$z$ quasars). A detailed analysis involving expected high-$z$ estimates of $H(z)$ from current planned BAO surveys is currently under investigation and will appear in a forthcoming communication.

\acknowledgments

JCC and JSA thank CNPq for the grants under which this work was carried out.


\begin{thebibliography}{99}

\bibitem{sand} A. R. Sandage, Physics Today {\bf{23}}, 3 (1970).

\bibitem{review}  T. Padmanabhan, Phys. Rept. {\bf{380}}, 235 (2003); J.~A.~Frieman, AIP Conf.\ Proc.\  {\bf 1057}, 87 (2008). arXiv:0904.1832; R. R. Caldwell and M. Kamionkowski, Ann.\ Rev.\ Nucl.\ Part.\ Sci.,   {\bf{59}}, 397 (2009).

\bibitem{steinhardt} I.~Maor, R.~Brustein and P.~J.~Steinhardt,  Phys.\ Rev.\ Lett.\  {\bf 86}, 6 (2001).  [Erratum-ibid.\  {\bf 87}, 049901 (2001)].

\bibitem{jimenez} R. Jimenez and A. Loeb, \apj {\bf{573}}, 37 (2002).

\bibitem{ma} T.~J.~Zhang and C.~Ma, ``Constraints on the Dark Side of the Universe and Observational Hubble Parameter Data,'' 
arXiv:1010.1307 [astro-ph.CO]; C.~Ma and T.~J.~Zhang, ApJ Lett. (in press). 
  arXiv:1007.3787 [astro-ph.CO].


\bibitem{pau} N.~Benitez {\it et al.}, ``Measuring Baryon Acoustic Oscillations along the line of sight with photometric redshifs: the PAU survey,''  astro-ph/0807.0535.

\bibitem{svj} J. Simon, L. Verde and R. Jimenez, \prd {\bf{71}}, 123001 (2005).


\bibitem{stern} D.~Stern, R.~Jimenez, L.~Verde, M.~Kamionkowski and S.~A.~Stanford,  JCAP {\bf 1002}, 008 (2010).

\bibitem{sdss7LRG} D.~P.~Carson and R.~C.~Nichol,
  Mon.\ Not.\ Roy.\ Astron.\ Soc.\  {\bf 408}, 213 (2010)


\bibitem{crawford} S. M. Crawford {\it et al.}, MNRAS {\bf{406}}, 2569 (2010).

\bibitem{riess} A. G. Riess {\it et al.}, Astrophys. J. {\bf{699}}, 539 (2009).

\bibitem{wmap} E.~Komatsu {\it et al.}  [WMAP Collaboration],
  Astrophys.\ J.\ Suppl.\  {\bf 192}, 18 (2011).



\bibitem{lyman}  P.~McDonald and D.~Eisenstein,
  Phys.\ Rev.\  D {\bf 76}, 063009 (2007); M.~L.~Norman, P.~Paschos and R.~Harkness,
  J.\ Phys.\ Conf.\ Ser.\  {\bf 180}, 012021 (2009)



\end{thebibliography}
\end{document}